\documentclass[RNAAS]{aastex63}

\graphicspath{{./}{figures/}}

\newcommand{\muhz}{\mbox{$\mu$Hz}}

\newcommand{\kepler}{{\em Kepler\/}}
\newcommand{\tess}{{\em TESS\/}}

\usepackage{CJKutf8}

\begin{document}

\title{Dealing with large gaps in asteroseismic time series}

\correspondingauthor{Timothy R. Bedding}
\email{tim.bedding@sydney.edu.au}

\author[0000-0001-5222-4661]{Timothy R. Bedding}
\affiliation{Sydney Institute for Astronomy, School of Physics, University of Sydney NSW 2006, Australia}
\affiliation{Stellar Astrophysics Centre, Aarhus University, DK-8000 Aarhus C, Denmark}

\author[0000-0002-9037-0018]{Hans Kjeldsen}
\affiliation{Stellar Astrophysics Centre, Aarhus University, DK-8000 Aarhus C, Denmark}

\begin{abstract}
With long data sets available for asteroseismology from space missions, it is sometimes necessary to deal with time series that have large gaps.  This is becoming particularly relevant for \tess, which is revisiting many fields on the sky every two years. Because solar-like oscillators have finite mode lifetimes, it has become tempting to close large gaps by shifting time stamps. Using actual data from the \kepler\ Mission, we show that this results in artificial structures in the power spectrum that compromise the measurements of mode frequencies and linewidths.  
\end{abstract}



\section{Introduction}

Asteroseismology involves studying the oscillations of stars by analysing the power spectra of their flux or radial-velocity variations.  With long data sets now available from space missions such as BRITE, CoRoT, \kepler, K2 and \tess, it is sometimes necessary to deal with time series that have large gaps.  This is becoming particularly relevant for \tess, which is revisiting many fields on the sky every two years \citep{Ricker++2015}.  

The Fourier transform of a long time series results in a very large array.  This is because the power spectrum is calculated with a step size that must sample the frequency resolution, and this is inversely proportional to the total duration of the time series (including the gap).  Because solar-like oscillators have finite mode lifetimes, it has become tempting to close large gaps by shifting time stamps and we are aware of several cases in which this has been done, both in published works \citep[e.g.,][]{Hekker++2010, Nielsen++2022} and in papers in preparation.  The justification is that mode lifetimes are much shorter than the gap, so shifting segments should not have much effect on the power spectrum. However, we argue that even if the modes are completely incoherent between the two segments, an arbitrary shift will introduce artificial structures in Fourier space that will compromise the profile fitting.  We demonstrate this with a simple test on a star observed by \kepler\ (we have tested other stars using \tess\ data and found similar results).

\section{Analysis and Results}

We used the subgiant star KIC~11137075 (also known as `Zebedee'), whose \kepler\ light curve contains more than one year of short-cadence data (1-minute sampling) that show high signal-to-noise solar-like oscillations centered at about 1700\,\muhz\ \citep{tian++2015}. For this test we defined two segments, as shown in the top panel of Fig~\ref{fig:zebedee}, where the boundaries coincided with the short breaks each month during which data were downloaded from the spacecraft \citep{Haas++2010}. The segment lengths were 30.1\,d and 26.0\,d, and the gap was 412.1\,d.

The other panels in the figure show power spectra in regions 2.5\,\muhz\ wide that are centred on four of the strongest modes (the first two are $l=0$ modes and the others are $l=1$). For each mode, the vertical dashed line marks the frequency measured by \citet{tian++2015} from the full \kepler\ time series.  The thin blue line shows the power spectrum calculated from the two segments when using the correct time stamps, that is, with the gap left in place.  The rapid oscillations are due to the spectral window from the large gap.  If desired, this spectrum could be smoothed to lower resolution and then re-sampled to produce a smaller array.
Another way to achieve the same result is to compute the power spectra of the two segments separately (using exactly the same frequency resolutions) and take the average.  This is shown by the black line in each figure.  We see that it follows quite closely the upper envelope of the blue line, although not exactly because of differences in the mode amplitudes between the two segments. Such differences reflect the well-know stochastic nature of solar-like oscillations.

The red line in Fig.~\ref{fig:zebedee} shows the power spectrum of a time series formed by shifting the segment~2 so that it follows immediately after segment~1.  The peaks in this spectrum are narrower than in the average of two segments (black line) but this does not indicate a real improvement in frequency resolution.  Rather, it is a natural consequence of placing the two segments side-by-side.  Furthermore, the structures in the red curve have been artificially introduced by the shifting procedure, and they greatly reduce the precision with the mode frequencies (and linewidths) can be measured.

\section{Discussion and Conclusions}

\begin{figure*}
\begin{center}
\includegraphics[width=0.8\linewidth]{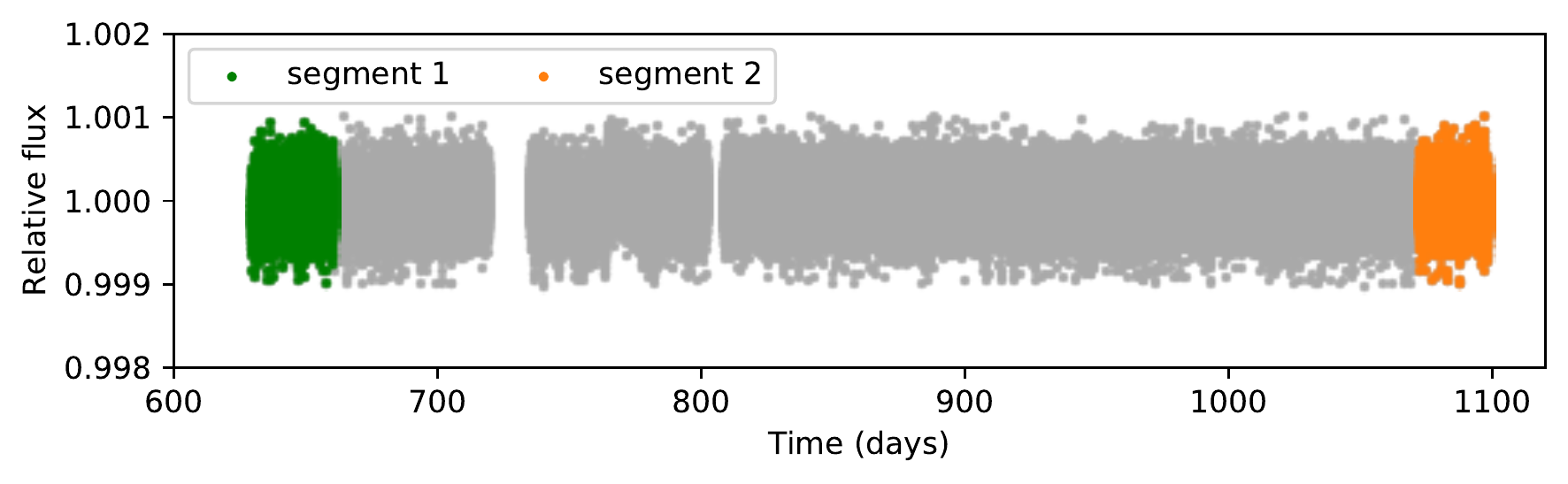}
\includegraphics[height=0.35\linewidth]{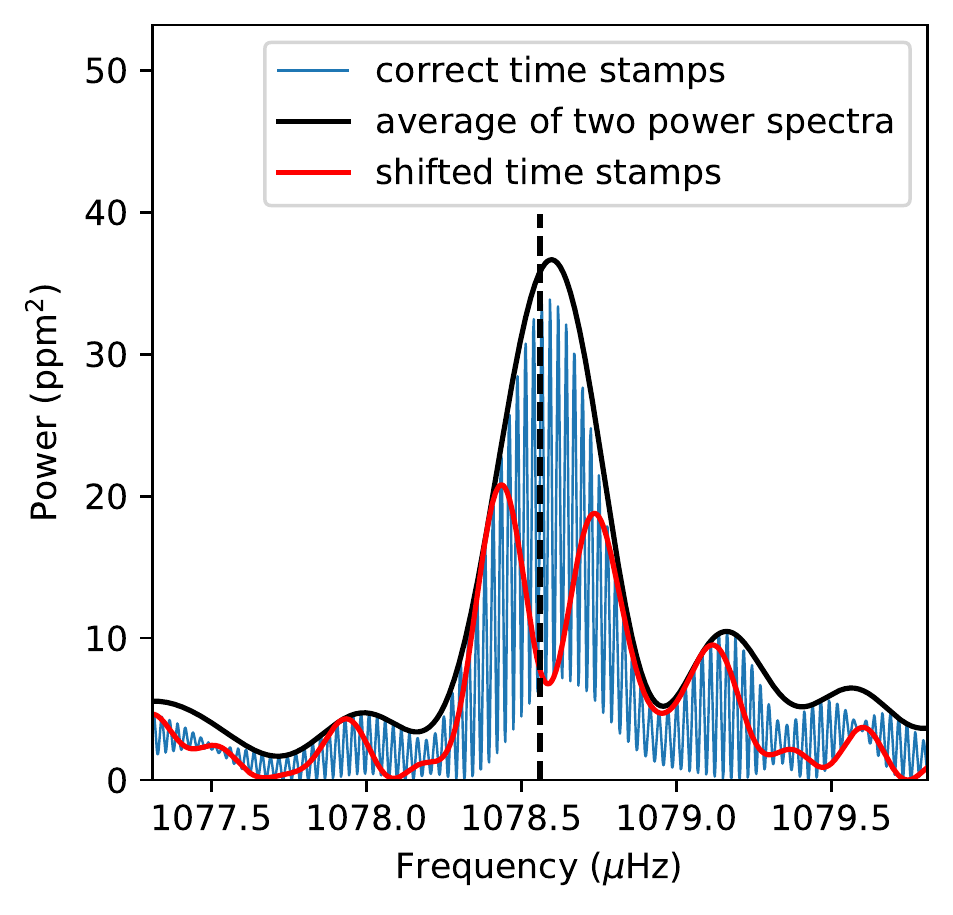}
\includegraphics[height=0.35\linewidth]{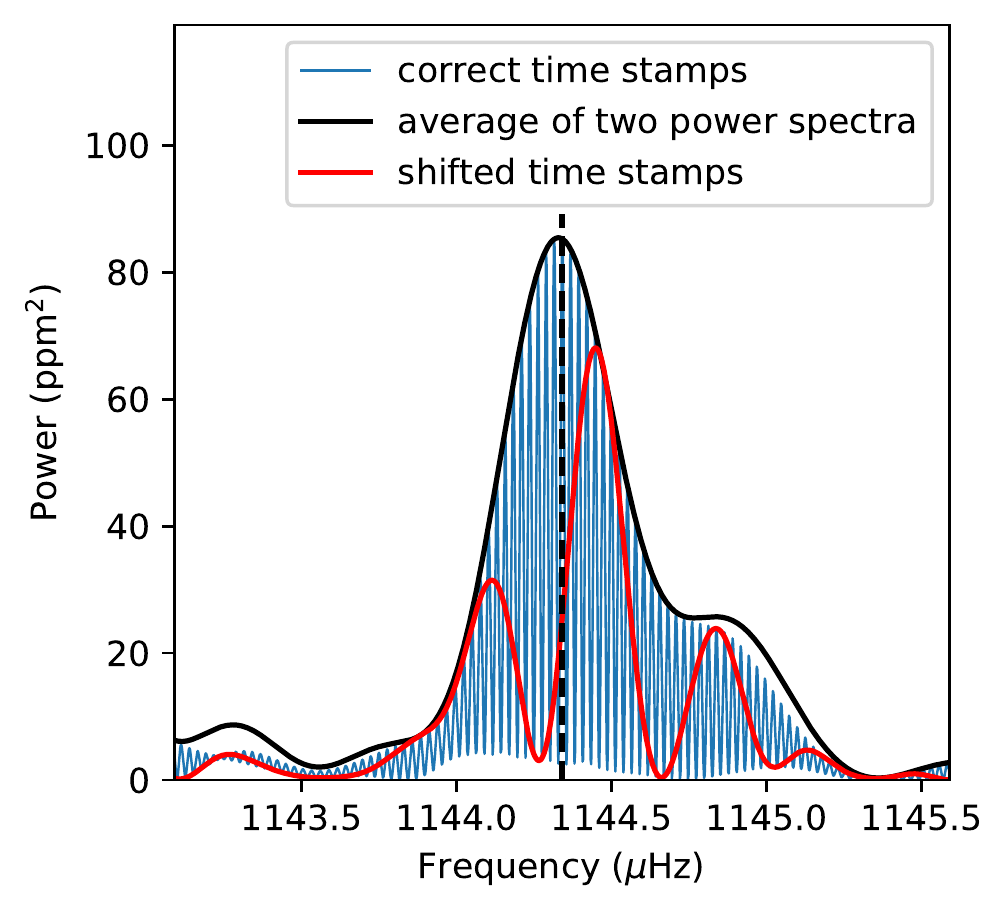}
\includegraphics[height=0.35\linewidth]{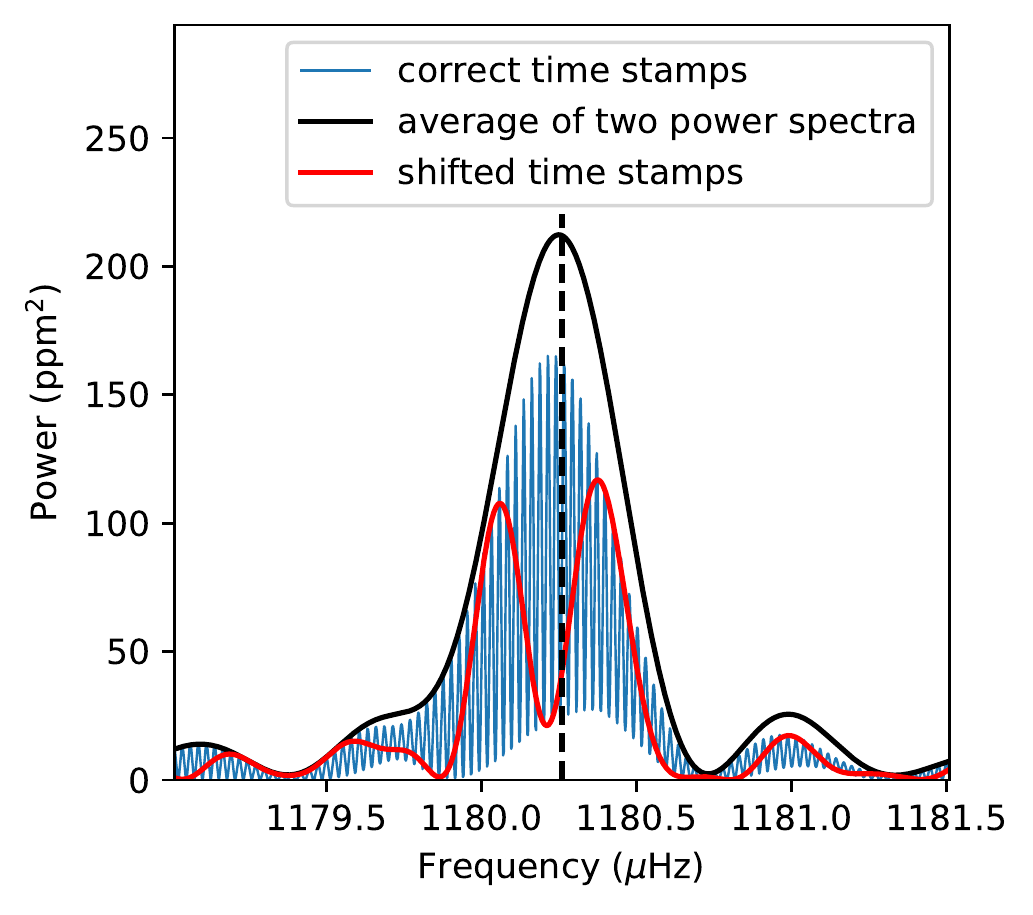}
\includegraphics[height=0.35\linewidth]{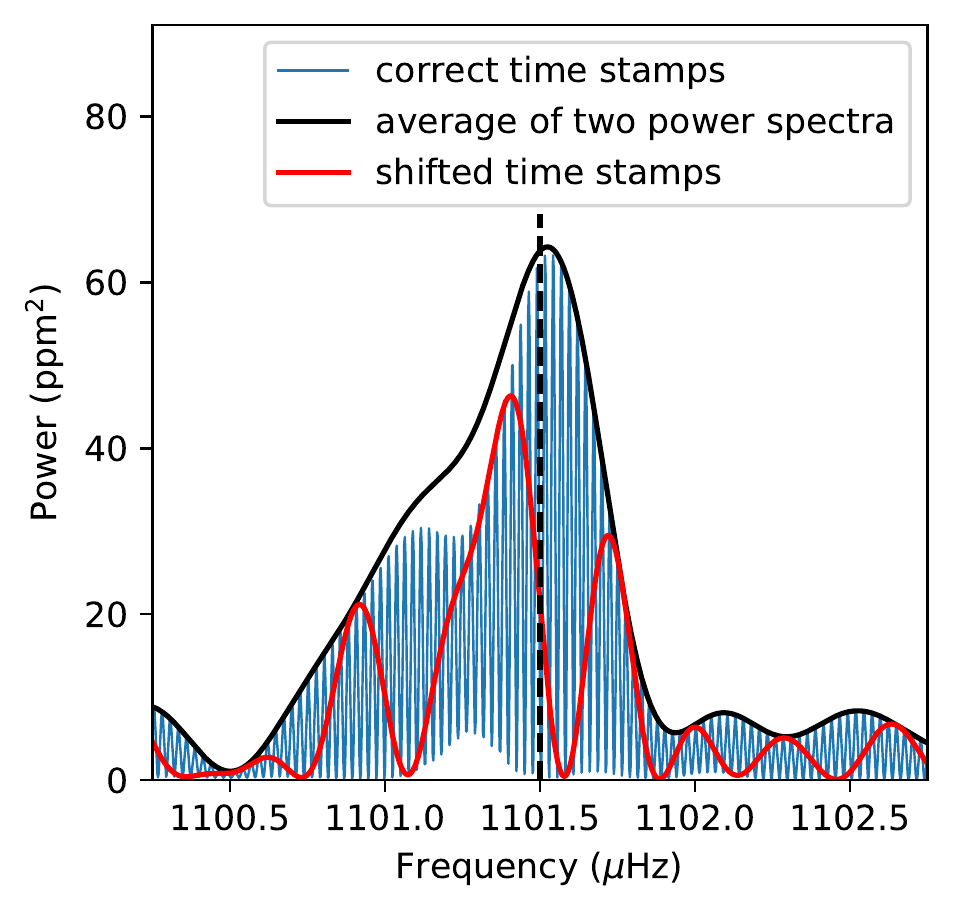}
\end{center}
\caption{The top panel shows 15 months of \kepler\ short-cadence data for KIC~11137075 (Quarters 7--11).  The other panels show the power spectra in narrow (2.5-\muhz) regions centred on four of the strongest modes (see text). The vertical dashed lines mark the frequencies measured from the full \kepler\ time series by \citet{tian++2015}. }.
\label{fig:zebedee}
\end{figure*}

To understand why the line profiles are modified by shifting time stamps, it helps to consider a simple scenario.  Suppose a pure sine wave with a single frequency is present in both segments, but with an arbitrary phase difference.  If the two segments are shifted so that they are adjacent, the power spectrum at the chosen frequency will depend very sensitively on the phase difference.  If the two signals happen to be in phase then the peak will be strong, but if they are out of phase then the peak will be very weak (and the power will be spread into nearby frequency bins).  Indeed, we have confirmed that the power spectrum in our test (red lines in Fig.~\ref{fig:zebedee}) changes greatly if the position of the shifted segment is slightly adjusted. Leaving a gap of only five minutes (about half the typical oscillation period) results in very different line profiles.

The procedure described above for averaging the individual power spectra of the segments (black lines in Fig.~\ref{fig:zebedee}) is only appropriate if the data segments have similar lengths.  In that case, segments with different noise levels---which  often occur with \tess\ data---can be combined by calculating a weighted average of the two power spectra.  In this case, the weights are easily calculated as the inverse of the square of the mean power level in a region of the power spectrum that is dominated by noise (usually at high frequencies).  If the segments do {\em not} have similar lengths then it is best to stay with the standard power spectrum, using the full data set, and to smooth to lower resolution if desired (see above).  In the data have non-uniform quality then the scatter in the time series can be used to calculate a {\em weighted} power spectrum (e.g., \citealt{Kjeldsen2005} and references therein). 
    
Finally, we always prefer to deal with power spectra that are over-sampled, typically by a factor of 5 or so.  This is because a critically-sampled spectrum does not contain the full information because it discards the phases, which means that genuine fine structure near the limit of the resolution is often not revealed.  


\acknowledgments

We thank the \kepler\ and \tess\ team for providing such wonderful data.  We gratefully acknowledge support from the Australian Research Council through Discovery Project DP210103119, and from the Danish National Research Foundation (Grant DNRF106) through its funding for the Stellar Astrophysics Centre (SAC). 

\facilities{Kepler,TESS}

\bigskip


\end{document}